# Dynamical *versus* statistical mesoscopic models for DNA denaturation

Marc JOYEUX and Ana-Maria FLORESCU

*Laboratoire de Spectrométrie Physique (CNRS UMR 5588),*

*Université Joseph Fourier - Grenoble 1,*

*BP 87, 38402 St Martin d'Hères, FRANCE*

**Abstract** : We recently proposed a dynamical mesoscopic model for DNA, which is based, like statistical ones, on site-dependent finite stacking and pairing enthalpies. In the present article, we first describe how the parameters of this model are varied to get predictions in better agreement with experimental results that were not addressed up to now, like mechanical unzipping, the evolution of the critical temperature with sequence length, and temperature resolution. We show that the model with the new parameters provides results that are in quantitative agreement with those obtained from statistical models. Investigation of the critical properties of the dynamical model suggests that DNA denaturation looks like a first-order phase transition in a broad temperature interval, but that there necessarily exists, very close to the critical temperature, a crossover to another regime. The exact nature of the melting dynamics in this second regime still has to be elucidated. We finally point out that the descriptions of the physics of the melting transition inferred from statistical and dynamical models are not completely identical and discuss the relevance of our model from the biological point of view.

email : Marc.JOYEUX@ujf-grenoble.fr

# 1 - Introduction

The unit that is commonly used to quantize the size of a DNA sequence is the "base pair" (bp). A "base pair" actually consists of the two A, T, G or C paired bases plus the two sugars and the two phosphate groups to which the bases are attached. Each base pair is approximately composed of 70 atoms, so that more than 200 coordinates are needed to specify its position in space and more than 400 ones to describe its dynamics. The obvious consequence is that only very short sequences and/or very fast phenomena can be modelled at the atomic level. For longer sequences and/or slower phenomena, one must have recourse to mesoscopic models, where groups of atoms are represented by a single particle in order to reduce as much as possible the number of degrees of freedom. This is precisely the case for the phenomenon this article deals with, namely the thermal denaturation of DNA, *i.e.* the separation of the two strands upon heating [1-6]. Examination of UV absorption spectra of diluted DNA solutions revealed a quarter of a century ago that the denaturation of inhomogeneous DNA sequences occurs through a series of steps as a function of temperature [7]. At each step, a large portion of the sequence (from hundred to several thousands bp) separates over a very narrow temperature interval, as in a first order phase transition. Various mesoscopic models have been developed along the years to investigate this point. Since modelling of denaturation requires that the evolution of several thousands bp be scrutinized, all these models agree in describing the state of each base pair with a very limited number of degrees of freedom - a single one in most cases. These models can be separated into two categories, namely statistical models and dynamical ones.

Statistical models describe a base pair of DNA like a 1/2 spin of the 1D Ising model, in the sense that it can only assume one of the two states "open" and "closed" [8,9]. A free energy change, which incorporates the effects of both pairing (intra base pair) and stacking (inter base pair) interactions, is associated with the opening of each base pair. Poland and Scheraga pointed out that such nearest-neighbour interactions are however not sufficient to induce a genuine phase transition (melting just corresponds to a smooth crossover between the closed helix form and the open coil state), but that the addition of an entropic term, which takes into account the number of configurations of the denaturated portions of the sequence (loops), induces long-range interactions that are weak but sufficient for a phase transition to occur [10,11]. Poland subsequently provided an algorithm which allows for the efficient calculation of the probability for each base pair to be in the open or closed state when considering both nearest-neighbour interactions and loop entropy [12]. This algorithm works particularly well when combined with the approximation proposed by Fixman and Freiere, which consists in expanding the loop function as an exponential series [13]. Such models were more recently further improved along two directions. First, the various parameters of the model were adjusted against experimental melting curves (see for example [14-16] and references therein). As a consequence, there now exist several online programs that provide within a few seconds reliable melting curves of sequences with several thousands bp [17], which is of great interest in many areas of biology, like PCR control and mutation analysis. Moreover, while the loop entropy was originally estimated by counting the number of configurations for a closed self-avoiding walk [18], it was shown very recently how to take more properly into account excluded volume effects between the loops and the rest of the chain [19,20]. This turns out to be of great importance from the physical point of view, since these later calculations lead to a loop closure exponent $c$ greater than 2, which implies that the phase transition is first order, while the older estimate of $c$ was smaller than 2 and therefore consistent with a second order phase transition.

As most colleagues, we use the statistical models just discussed to get rapid and reliable estimations of the denaturation properties of precise DNA sequences. Yet, we believe that dynamical models provide an interesting complementary point of view on melting. Dynamical models are based on explicit expressions of the energy of the DNA sequence, which are written in terms of usual continuous coordinates and velocities. Dynamical models are conceptually appealing in the sense that one just needs to provide a microscopic description of the system, like kinetic energy and the shape of pairing and stacking interactions. Its macroscopic properties and evolution with temperature then unequivocally follow there from. Stated in other words, if the masses and characteristic energies introduced in the Hamiltonian are reasonable and the derived macroscopic properties match experimental results, then one might feel confident that the microscopic description of the system is sound. Moreover, dynamical models are of course mandatory as soon as one is not interested in averaged quantities but rather in transient phenomena and fluctuations [21].

To our knowledge, Prohofsky and co-workers were the first to propose a dynamical model capable of describing DNA denaturation [22,23]. In this model, the pairing interaction is taken in the form of a Morse potential expressed as a function of the distance between paired bases, while the stacking interaction is considered to be harmonic with respect to the distance between successive bases on the same strand. Dauxois, Peyrard and Bishop (DPB) pointed out that this model leads to a much too smooth denaturation process, but that this is no longer the case upon introduction of a weaker stiffness for single stranded DNA compared to double-stranded DNA [24]. In contrast with the model of Prohofsky and co-workers, the stacking interaction in the DPB model is not harmonic, but it still differs fundamentally from that of statistical models because it does not make reference to any characteristic energy. Since its introduction, this model was used to unravel several aspects of melting, see for example [6] and references therein. Note, however, that calculations performed with dynamical models are many orders of magnitude slower than statistical ones, so that the DPB model was essentially used to investigate the dynamics of short sequences containing from several tens to a few hundreds bp [25,26]. Moreover, no attempt was made to check whether this model is able to reproduce the characteristic peaks which appear in the melting curves of inhomogeneous sequences in the 1000-10000 bp range (the peaks that were reported for periodic DNA sequences with two or three bp in the unit cell [27] essentially arise from end effects and are not directly related to the experimentally observed ones).

We proposed a couple of years ago a dynamical model for DNA [28], which is closer to statistical ones than the DPB model in the sense that it is based on site specific stacking enthalpies (numerical values for the enthalpies were borrowed from table 1 of [17]), and showed that the finiteness of the stacking interaction is in itself sufficient to insure a sharp melting transition. We next performed a complete statistical analysis of DNA melting, including finite and inhomogeneous sequences, on the basis of this model [29-31], and concluded that the transition looks first order over a large temperature interval, in agreement with the DPB model. We furthermore observed that the denaturation curves of inhomogeneous sequences in the 1000-10000 bp range obtained with this type of models indeed display the characteristic peaks denoting sequential opening of the sequence [31,32], although temperature resolution appears to be somewhat lower than for statistical models and experiments. Additional calculations convinced us that our model can also be improved with respect to at least two other points. First, our model, like the DPB one, predicts a much too large sensitivity of the critical temperature with respect to the length of the sequence [30], leading to unrealistically small melting temperatures for sequences with less than several hundreds bp. Moreover, we performed some calculations to probe the mechanical unzipping of DNA and found that our model predicts a too small critical force at ambient temperature. It turns out that the description of these three points - temperature resolution, influence of the sequence length on critical temperature, and critical force in mechanical unzipping - can be greatly improved without modifying the general form of the Hamiltonian, that is, just by slightly varying the numerical values of the parameters of the model. The model contains few parameters, but the experimental melting curves that were used to adjust them are not sufficient to fix all of them unambiguously. Stated in other words, the temperature resolution, the dependence of the critical temperature with respect to the sequence length, and the critical force of mechanical unzipping experiments can be considered as additional experimental results against which the parameters of the model can be adjusted.

The purpose of this article is thus threefold. We first describe (in section 2) how the parameters of the model are varied to get predictions in better agreement with the additional experimental results, before showing (in section 3) that the model with the new parameters provides results that are in quantitative agreement with those obtained from statistical models. We next investigate briefly (in section 4) the critical properties of the new model and show that they do not differ significantly from that of the former model. We however discuss in more detail than earlier the critical behaviour in the very narrow region just below the critical temperature. We finally point out in section 5 that, although the melting curves obtained with the new dynamical model and statistical ones are very similar, the physical descriptions of the melting transition inferred from the two types of models are not identical. Moreover, we discuss in the same section the importance of our study from the biological point of view.

**2 - Adjustment of the parameters of the dynamical model**

*2.1 - Expression of the Hamiltonian*

The general form of the model we are considering in this paper is

$$H = T + V$$

$$T = \frac{m}{2} \sum_{n=1}^{N} \left(\frac{dy_n}{dt}\right)^2$$

$$V = \sum_{n=1}^{N} V_M^{(n)}(y_n) + \sum_{n=2}^{N} W^{(n)}(y_n, y_{n-1}) \quad (2.1)$$

$$V_M^{(n)}(y_n) = D_n (1 - \exp(-a y_n))^2$$

$$W^{(n)}(y_n, y_{n-1}) = \frac{\Delta H_n}{C} \left(1 - \exp(-b(y_n - y_{n-1})^2)\right) + K_b (y_n - y_{n-1})^2$$

where $N$ is the number of base pairs in the sequence and $y_n$ a measure of the distance between the paired bases at position $n$. More precisely, if $u_n$ and $v_n$ denote the displacements of the two bases of pair $n$ from their equilibrium positions along the direction of the hydrogen bonds that connect the two bases in the pair, then $y_n$ is defined as $y_n = (u_n - v_n)/\sqrt{2}$ (see for example equation (4) of [23] or the equation at the bottom of page 2755 of [33]). In the expression for the kinetic energy $T$, $m$ denotes the mass of a nucleotide, which we assume to be independent of the precise nature of the base pair at position $n$ (numerically, we use $m$=300 amu). The potential energy $V$ is the sum of two different contributions, namely on-site potentials $V_M^{(n)}(y_n)$ and nearest-neighbour interaction potentials $W^{(n)}(y_n, y_{n-1})$. $V_M^{(n)}(y_n)$ represents the two or three hydrogen bonds that connect the paired bases at position $n$ and is taken as a Morse potential of depth $D_n$. $V_M^{(n)}(y_n)$ is often called a "pairing" potential, because it is an increasing function of the distance between the two bases of pair $n$, $\sqrt{2}|y_n|$, and therefore opposes the dissociation of the pair. $W^{(n)}(y_n, y_{n-1})$ is again the sum of two terms, namely the stacking potential plus the backbone stiffness. Both terms are increasing functions of $|y_n - y_{n-1}|$, which means that they oppose the de-stacking of the bases, i.e. the separation of successive bases belonging to the same strand. The stacking potential essentially results from hydrophobic interactions with the solvent and electronic interactions between successive base pairs on the same strand and is modelled by a gaussian hole of depth $\Delta H_n / C$. At last, the backbone stiffness is taken as a harmonic potential of constant $K_b$. Its role consists in preventing dislocation of the strands, that is, in insuring that base pairs belonging to the same strand do not separate infinitely when approaching the melting temperature.

The numerical values for the parameters of (2.1) we used in previous works [21,28-32] were obtained in the following way. The ten stacking enthalpies $\Delta H_n$ were borrowed from statistical models (table 1 of [17]) and we assumed that the paired bases do not unstack simultaneously, which implies that $C = 2$ [28]. $D_n = 0.04$ eV and $a=4.45$ Å$^{-1}$ were taken from the DPB model [24], while $K_b = 10^{-5}$ eV Å$^{-2}$ was fixed somewhat arbitrarily. We finally varied $b$ and found that the choice $b=0.10$ Å$^{-2}$ leads to sharp denaturation curves, to a separation of about 40 to 50 K between the critical temperatures of pure AT and pure GC long sequences, and to some temperature pattern in the denaturation curve of inhomogeneous sequences. On the other hand, we assumed a uniform stacking strength $\Delta H_n / C = 0.22$ eV to model the homogeneous sequences that are involved in most statistical studies.

As briefly mentioned in the Introduction, the results obtained with this set of parameters can however be improved with respect to at least three points. First, the denaturation curves obtained with these parameters lack some temperature resolution. For example, the curve shown in figure 5 of [31] is somewhat too smooth compared to the curves obtained with statistical models (see figure 2 of [34]) or experimental ones. Moreover, the critical temperature diminishes much too quickly with decreasing sequence lengths. As reported in [30], the lowering of the melting temperature behaves as 3250/$N$ for our model and 1850/$N$ for the DPB model, while most online oligonucleotide property calculators assume a 500/$N$ dependence (which agrees with the experimental results reported in [35]) and statistical models even predict a gap smaller than 1 K between the melting temperatures of an infinitely long homogeneous sequence and its finite counterpart with $N$=100 bp. Finally, mechanical unzipping experiments performed at constant force show that the critical force, which is needed to open a DNA sequence around 20°C, lies in the range 10-20 pN [36,37], while our model predicts that a few pN are sufficient (see below). As will be argued in the remainder of this section, this poor agreement between predicted and measured critical forces is essentially ascribable to a too small value of $K_b$ (remember that this parameter was fixed somewhat arbitrarily), while the exaggerated sensitivity of the melting temperature with sequence length results from the too large depth of the stacking interaction. We will see that it is sufficient, once these two points have been corrected, to adjust slightly the remaining parameters in order to reproduce correctly experimental denaturation curves. The good surprise actually comes from the fact that the resolution in temperature is then also substantially higher and close to the experimental one.

*2.2 - Taking mechanical unzipping into account*

In the presence of a force acting on one of the base pairs at position $n=1$, the energy $H_{stretch}$ of the system may be written as [38]

$$H_{stretch} = H - F y_1 \qquad (2.2)$$

It should however be noted that, because of the $\sqrt{2}$ factor that appears in the expression of $y_n$ as a function of the positions $u_n$ and $v_n$ of each nucleotide (see below equation (2.1)), $F$ is, properly speaking, not the experimental force, but rather the experimental force multiplied by $\sqrt{2}$. According to most authors, the critical force $F_c(T)$ required to unzip double-stranded DNA at temperature $T$ is the force for which the variation of the average free energy per base pair of a very long sequence, $g_0(T)$, is equal to the variation of free energy per base pair of the stretched single strands, $g_u(T,F)$ [36-40]

$$g_u(T, F_c(T)) - g_u(T_c, 0) = g_0(T) - g_0(T_c) \tag{2.3}$$

Equation (2.3) contains an approximation, in the sense that it assumes that the free energy per base pair of a stretched double-stranded sequence is equal to that of an unstretched sequence. Results obtained using equation (2.3) are therefore better checked with independent calculations. Following the arguments of Singh and Singh [38], the variation of the free energy per base pair of unstretched long sequences may be estimated for the model in (2.1) according to

$$g_0(T) - g_0(T_c) = D\left(\frac{T}{T_c} - 1\right) \tag{2.4}$$

for temperatures $T$ smaller than the critical temperature $T_c$. In (2.4) $D = D_n$ denotes the depth of the Morse potential for a homogeneous sequence. Moreover, it is possible for some models to calculate the free energy per base pair of the stretched single strands by taking the derivative of the partition function $Z(T,F)$

$$Z(T,F) = \int dy_1 \, dy_2 \ldots dy_N \, \exp(-\beta H_{stretch})$$
$$= \int dy_1 \, dy_2 \ldots dy_N \, \exp\left(-\beta\left(ND + \sum_{n=2}^{N}\{W^{(n)}(y_n, y_{n-1}) + F(y_n - y_{n-1})\}\right)\right) \tag{2.5}$$

$$g_u(T,F) = -\frac{1}{N\beta}\ln(Z(T,F))$$

where $\beta = 1/(k_B T)$. When approximating the nearest-neighbour interaction potential in (2.1) by

$$W^{(n)}(y_n, y_{n-1}) \approx \min\left[\frac{\Delta H}{C}, \frac{\Delta H b}{C}(y_n - y_{n-1})^2\right] + K_b(y_n - y_{n-1})^2 \tag{2.6}$$

one obtains

$$g_u(T,F) = D - \frac{1}{\beta}\ln(I_1 + I_2) \tag{2.7}$$

where

$$I_1 = \sqrt{\frac{\pi}{4\beta K_b}} \exp\left(u^2 - \frac{\beta \Delta H}{C}\right)\{2 - \text{erf}(v-u) - \text{erf}(v+u)\}$$

$$u = \sqrt{\frac{\beta F^2}{4K_b}} \tag{2.8}$$

$$v = \sqrt{\frac{\beta K_b}{b}}$$

and

$$I_2 = \sqrt{\frac{\pi}{4\beta\kappa}} \exp(u'^2) \{\text{erf}(v'-u') + \text{erf}(v'+u')\}$$

$$\kappa = K_b + \frac{\Delta H\, b}{C}$$

$$u' = \sqrt{\frac{\beta F^2}{4\kappa}}$$

$$v' = \sqrt{\frac{\beta\kappa}{b}}$$

(2.9)

The plot of $F_c(T)$ obtained with equations (2.3), (2.4) and (2.7) and the parameters reported in section (2.1) is shown as a solid line in figure 1. We checked that Monte-Carlo simulations performed with the Hamiltonian in (2.2), reported as open circles in figure 1, are in excellent agreement with this curve in the 290-370 K temperature range. Conclusion therefore is, that the parameters used up to now lead to a much too small critical force around 20°C (the experimental value lies in the range 10-20 pN [36,37]), especially when remembering that $F$ in equations (2.2) to (2.9) denotes the experimental force multiplied by $\sqrt{2}$. Examination of equations (2.7)-(2.9) shows that $F_c(T)$ depends strongly on $K_b$, which was fixed somewhat arbitrarily in the first set of parameters. Comparison with mechanical unzipping experiments will therefore help fix this parameter to a more grounded value. Equation (2.7) indicates that $K_b$ must actually be increased for the calculated critical curve to come closer to the experimental one. This is a very positive point, because we also noticed that $K_b = 10^{-5}$ eV Å$^{-2}$ sometimes leads to distances between successive bases on the same strand that are unrealistically large. Increasing $K_b$ will therefore improve the quality of the model with respect to two points and not only one.

*2.3 - Taking sequence length effects into account*

Although we have no definitive proof thereof, many trials convinced us that the only way to reduce substantially the dependence of the critical temperature with respect to the length of the sequence consists in decreasing the depth of the stacking interaction, that is, in assuming smaller values for $\Delta H_n/C$. Since we want to go on using the stacking enthalpies $\Delta H_n$ that were adjusted for statistical models (like those in table 1 of [17]), $C$ must consequently be made larger than the value $C=2$, which we used up to now and was obtained by assuming that paired bases do not unstack simultaneously [28]. Low frequency Raman spectra [41,42] and theoretical investigations of collective modes in DNA [43,44] suggest, on the other hand, that the stacking stiffness $\Delta H_n b/C$ may be larger than what was assumed in our original set of parameters (note, however, that the two models are substantially different). The increase of the parameter $b$ controlling the width of the gaussian hole must therefore be larger than the decrease of the hole depth $\Delta H_n/C$.

*2.4 – New parameters*

Taking into account the considerations in sections (2.2) and (2.3) and requiring that calculated melting curves reproduce experimental ones is still not sufficient to fix all the parameters of the model unambiguously. However, we found that the following set of parameters allows for a correct reproduction of most of the known properties of DNA under usual salinity conditions (75 mM NaCl) :

(*) Morse pairing potential : $D_n = D = 0.048$ eV and $a = 6.0$ Å$^{-1}$ (against $D = 0.040$ eV and $a = 4.45$ Å$^{-1}$ previously).

(*) Stacking potential : $C = 4$ and $b = 0.80$ Å$^{-2}$ (against $C = 2$ and $b = 0.10$ Å$^{-2}$ previously). For inhomogeneous sequences, the ten stacking enthalpies $\Delta H_n$ are taken from table 1 of [17] (as previously), while $\Delta H_n = \Delta H = 0.409$ eV for homogeneous ones (against $\Delta H = 0.44$ eV previously).

(*) Backbone stiffness : $K_b = 4.0 \ 10^{-4}$ eV Å$^{-2}$ (against $K_b = 10^{-5}$ eV Å$^{-2}$ previously). As anticipated, the by far most important change compared to the earlier set of parameters concerns this parameter.

The plot of $F_c(T)$ obtained with the new set of parameters and equations (2.3), (2.4) and (2.7) is shown in figure 1. It is seen that, although probably still somewhat too small, the critical force around 20°C is now in much better agreement with the experimentally determined one [36,37]. Moreover, the melting temperature also decreases much less rapidly as a function of the length of the sequence. Figure 2 indeed indicates a $540/N^{0.77}$ dependence, which is still larger than the decrease predicted by statistical models but is in qualitative agreement with both the $500/N$ dependence that is plugged in most online oligonucleotide property calculators and the experimental results reported in [35].

Additional results obtained with this new set of parameters will be presented and discussed in sections 3 and 4.

*2.5 – Heterogeneous pairing and salt concentration contribution*

The set of parameters proposed above assumes that heterogeneity is carried by stacking interactions. One might instead assume that heterogeneity is carried by pairing interactions, as in heterogeneous versions of the DPB potential [25-27]. It is sufficient, for this purpose, to fix $\Delta H_n$ to its average value $\Delta H = 0.409$ eV and introduce two different values for the Morse potential depth $D_n$, namely one for AT base pairs and one for GC base pairs. One is thus led to the following set of parameters :

(*) Morse pairing potential : $D_n = 0.041$ eV for AT base pairs, $D_n = 0.054$ eV for GC base pairs, and $a = 6.0$ Å$^{-1}$.

(*) Stacking potential : $C = 4$, $b = 0.80$ Å$^{-2}$ and $\Delta H_n = \Delta H = 0.409$ eV.

(*) Backbone stiffness : $K_b = 4.0 \ 10^{-4}$ eV Å$^{-2}$.

Results obtained with this set of parameters are qualitatively and quantitatively similar to those obtained with the set in section (2.4). Since recent work suggests that heterogeneity is carried by both pairing and stacking interactions [45-47], one could think of introducing both different $D_n$ and $\Delta H_n$ values in the model. We however made no attempt in this direction because of the complexity of Transfer-Integral calculations for this kind of hybrid models [31].

At last, it should be noted that the influence of different salinity conditions on DNA melting can easily be taken into account in this particular form of our model by using $D_n = 0.041 + 0.006 \, \text{Log}([\text{Na}^+]/[\text{Na}^+]_0)$ for AT base pairs and $D_n = 0.054 + 0.004 \, \text{Log}([\text{Na}^+]/[\text{Na}^+]_0)$ for GC base pairs, where $D_n$ is expressed in eV and $[\text{Na}^+]_0 = 75$ mM. The variations of critical temperature with respect to salinity obtained with these expressions agree well with those predicted by statistical models.

**3 – Comparison of melting curves obtained with statistical and dynamical models**

In this section, we first provide a brief outlook of the two methods we used to investigate the properties of the model described in sections (2.1) and (2.4) - the interested reader is referred to our previous work [21,28-32] for more detail -, before showing that its melting properties are very close to those of statistical models.

*3.1 – Molecular Dynamics simulations and Transfer Integral calculations*

The two methods we used to investigate the properties of the model described in sections (2.1) and (2.4) are Molecular Dynamics (MD) simulations and Transfer-Integral (TI) calculations. MD simulations consist in integrating step by step Langevin equations

$$m \frac{d^2 y_n}{dt^2} = -\frac{\partial H}{\partial y_n} - m\gamma \frac{d y_n}{dt} + w(t)\sqrt{2m\gamma k_B T} \qquad (3.1)$$

with a second order Brünger-Brooks-Karplus integrator [48]. γ is the dissipation coefficient (we assumed $\gamma = 5$ ns$^{-1}$) and $w(t)$ a normally distributed random function with zero mean and unit variance. The second and third term in the right-hand side of (3.1) model the effect of the solvent on the DNA sequence. The sequence is first heated by submitting it to a temperature ramp, which is slow enough for the physical temperature of the system (calculated from the average kinetic energy) to follow the temperature of the random kicks (the symbol $T$ in (3.1)). The average values of the quantities we are interested in are then obtained by integrating Langevin equations at constant temperature for time intervals of 100 ns. MD simulations are very easy to implement, but they have two limitations. First, they require a very large amount of CPU time, because step by step integration of hundreds or thousands of coupled differential equations is intrinsically slow. Moreover, temperature resolution is rather poor, especially close to the melting temperature, because of the very slow fluctuations of temperature in this range [21].

On the other side, the TI method [27,49] is a mathematical technique to replace the $N$-dimensional integrals, which appear for example in the expressions of the partition function $Z$

$$Z = \int dy_1\, dy_2 \ldots dy_N \exp(-\beta V) \tag{3.2}$$

and the average bond length at position $n$

$$\langle y_n \rangle = \frac{1}{Z} \int dy_1\, dy_2 \ldots dy_N\, y_n \exp(-\beta V), \tag{3.3}$$

by products of $N$ one-dimensional integrals. Other quantities of interest, like the free energy per base pair, $f$, the entropy per base pair, $s$, and the specific heat per base pair, $c_V$, are easily obtained from $Z$ according to

$$\begin{aligned} f &= -\frac{1}{N\beta} \ln(Z) \\ s &= -\frac{\partial f}{\partial T} \\ c_V &= -T \frac{\partial^2 f}{\partial T^2} \end{aligned} \tag{3.4}$$

Note that we use finite differences for the calculation of $s$ and $c_V$. *When it works*, the TI method is very efficient, in the sense that it enables to calculate most quantities much more rapidly and with a better temperature resolution than MD simulations. As discussed in some detail by Zhang *et al* [27], the TI kernel is however singular when using a bound on-site Morse potential, so that one needs to check carefully the convergence of the obtained results with respect to the upper bound for $y$ which is assumed in practical calculations. Our general observation is that, at the thermodynamic limit of infinitely long homogeneous chains, there always exists a certain temperature range surrounding the critical temperature, where the TI method is not valid. For some sets of parameters, this interval is so large that the TI method is essentially useless (this is, of course, not the case for the set of parameters proposed in section (2.4), see section 4). In contrast, calculations are more reliable for finite sequences, because they melt at temperatures that are lower than the critical temperature of the infinite sequence.

In order to illustrate the capabilities of the two methods, we compare in figures 3 and 4 the temperature evolution of the average base pair separation $\langle y \rangle = \frac{1}{N} \sum \langle y_n \rangle$ (figure 3), as well as the average energy in each Morse potential and the average total potential energy per base pair $u = \langle V \rangle / N$ (figure 4), obtained with TI calculations (solid lines) and MD simulations (dashed lines) for an homogeneous sequence with 1000 base pairs. It is seen that the agreement between both types of calculations is generally very good, except close to the critical temperature, where MD simulations are much noisier than TI calculations (although 10 trajectories were averaged, so that MD simulations required between 10 and 100 times the CPU time of TI calculations) and evolve less sharply with temperature. As mentioned above, this difference is due in part to the very slow temperature fluctuations of the sequence in this interval [21] and in part to the fact that the averaging time between two temperature increments (100 ns per K) is too small compared to the characteristic times of the denaturation dynamics of the sequence.

*3.2 – Comparison of the melting curves obtained with this model and statistical ones*

As mentioned in the Introduction, a base pair can only assume one of the two states "open" or "closed" in statistical models. There is therefore no ambiguity when estimating, for example, the fraction of open base pairs as a function of temperature, or the temperature at which each base pair of a given sequence opens. Such plots are shown as dashed lines in figures 5 and 6 for the 1793 bp human β-actin cDNA sequence (NCB entry code NM_001101). Calculations were performed with the MELTSIM program [17], the parameters of Blossey and Carlon [16], and a salinity $[Na^+]_0 = 75$ mM.

"Closed" and "open" are more ambiguous concepts in the case of dynamical models, which are expressed in terms of continuous coordinates $y_n$. For example, one might consider that the fraction of open base pairs is obtained by computing at each time $t$ the fraction of base pairs for which $y_n$ is larger than a given threshold $y_{thresh}$ and in subsequently averaging this quantity over $t$ [24-27,33]. Alternatively, one can consider that a given base pair $n$ is open if the mean elongation $\langle y_n \rangle$ is larger than the threshold $y_{thresh}$ and average this quantity over the sequence. The two definitions are rather close and, as long as one does not deal with experimental results obtained with ultra-short laser pulses, there is no physical reason to choose one definition instead of the other. Still, the curves obtained with these two definitions are not identical. In particular, we noticed that results obtained with the second definition are better resolved in temperature and closer to those obtained with statistical models [31,32]. In the remainder of this paper, we will therefore use this definition and consider that base pair $n$ is open if $\langle y_n \rangle > y_{thresh}$. It remains that the choice of $y_{thresh}$ itself is not trivial. Figure 3 indeed shows that if one chooses for $y_{thresh}$ a too small value, like for example two or three times $1/a$ (approximately 0.5 Å), then application of the criterion to a long homogenous sequence would lead to the erroneous conclusion that all base pairs are already open tens of Kelvins below the critical temperature. For such long homogeneous sequences, the larger the value of $y_{thresh}$, the closer the critical temperature determined with the $\langle y_n \rangle > y_{thresh}$ criterion to the exact one. But, on the other hand, a too large value of $y_{thresh}$ is in turn not suitable for inhomogeneous sequences, because different portions of an inhomogeneous sequence melt at different temperatures and the separation of open base pairs belonging to bubbles is limited by the double-stranded portions. The choice of $y_{thresh}$ therefore appears as a compromise between these two conflicting considerations.

Practically, we found that, for the model proposed in section 2, the choice $y_{thresh} = 10$ Å leads to reasonable results for both homogeneous and inhomogeneous sequences. Still, one must keep in mind that the critical temperature determined with this criterion is 2 to 3 Kelvins lower than the exact one (see figure 3).

We computed the evolution of the fraction of open base pairs as a function of temperature and the melting temperature of each base pair of the 1793 bp actin sequence for the model of section 2 using the TI procedure described in [31]. The results are plotted as solid lines in figures 5 and 6. The results obtained with three different thresholds ($y_{thresh} = 7$, 10 and 15 Å) are shown for the sake of comparison. It can be checked that, except for the short portion of the sequence that melts at the highest temperature, the agreement between results obtained with statistical and dynamical models is rather striking. In particular, the resolution in temperature of melting curves is higher for the new parameters than for the old set (compare figure 5 with figure 5 of [31]) and almost comparable to that of statistical models. In contrast, no increase in resolution is observed in the plot of $c_V(T)$ (compare figure 7 with figure 6 of [31]).

**4 – Critical behaviour of the dynamical model**

This section is devoted to the description of the critical behaviour of the dynamical model of section 2. We will show that it does not differ significantly from the behaviour observed with the old set of parameters [29,30], which implies that the melting of homogeneous DNA sequences looks like a first-order phase transition. We will however point out that there is necessarily a crossover to another regime very close to the melting temperature.

The temperature evolution of the entropy per base pair, *s*, is show in figure 8 for infinitely long sequences and sequences with *N*=1000 and *N*=100 bp. This plot, as well as all the other plots discussed in this section, were obtained from TI calculations performed as discussed in [29,30]. It is seen that the temperature evolution of *s* displays the step-like behavior, which is characteristic of first-order phase transitions. This is particularly clear for the infinite sequence and the sequence with *N*=1000 bp, but the step-like behavior is still well-marked for shorter sequences. As usual, this step-like behavior of *s* corresponds to thin peaks in the temperature evolution of the specific heat per base pair, $c_V$, as can be checked in figure 9. Note that, in both figures, the solid line associated with infinite sequences is interrupted in the narrow temperature interval where TI calculations are not valid.

Further information is gained by calculating the critical exponents, which characterize the power-law behavior of several statistical properties of infinitely long homogeneous sequences close to the critical temperature. For example, critical exponents α, β and ν are defined according to

$$c_V \propto (T_c - T)^{-\alpha}$$
$$\langle y \rangle \propto (T_c - T)^\beta \tag{4.1}$$
$$\xi \propto (T_c - T)^{-\nu}$$

where ξ denotes the correlation length and $\langle y \rangle$ is taken as the order parameter of the melting transition [29]. The critical temperature of a sequence of length $N$, $T_c(N)$, is easily found as the temperature where $c_V$ is maximum. Because of the temperature interval where TI calculations are not valid, it may be somewhat more complex to determine the critical temperature of infinitely long sequences, $T_c = T_c(N = \infty)$. In this work, we took advantage of the fact that the critical temperature shift $T_c - T_c(N)$ unambiguously decreases as a power of $N$: we consequently found $T_c$ as the temperature for which $\log(T_c - T_c(N))$ is best adjusted with a linear function of $\log(N)$. One gets $T_c = 359.43$ K and, as already mentioned in section 2.4, $T_c(N) \approx T_c - 539.5 N^{-0.770}$ (see figure 2). Critical exponents α, β and ν are then obtained by drawing log-log plots of, respectively, $c_V$, $\langle y \rangle$ and ξ as a function of the temperature gap $T_c - T$ and in estimating the slope of each curve in the temperature range where the power law holds. The plots in figures 10 to 12 show that $\alpha = 1.33$, $\beta = -1.41$, and $\nu = 1.47$, not so far from the values $\alpha = 1.13$, $\beta = -1.31$, and $\nu = 1.23$ obtained with the old set of parameters [29].

At that point, three comments are worthwhile. First, the critical exponent of the specific heat, α, is larger than 1, which confirms that melting indeed looks like a first-order phase transition in the temperature range where power-laws hold. Moreover, not all scaling laws are satisfied for the model of section 2.4. This is a conclusion we already arrived at for the DPB model and our model with the old set of parameters [29]. More precisely, Rushbrooke and Widom identities, which are based on the homogeneity assumption, are satisfied, while Josephson and Fisher identities, which rely on the more compelling generalized homogeneity assumption, are not. For example, one can check from the numerical values above that $2 - \alpha < \nu d$. In [29], we tentatively assigned this failure of Josephson identity to the divergence of the average bond length, which may invalidate the assumption that the correlation length is solely responsible for singular contributions to thermodynamic quantities. At last, it can be seen in figures 10 and 11 that the critical behaviour of the homogeneous sequence with 1000 bp is close to that of the infinitely long sequence. As argued in [31], this indicates that, within the validity of our model, one may indeed describe the denaturation of long inhomogeneous sequences as a succession of first-order phase transitions.

Pr Miguel A. Munoz (Universidad de Granada, Spain) however attracted our attention to the fact that the first-order regime with $\alpha > 1$ cannot hold up to the critical temperature, because the average potential energy per base pair, $u = \langle V \rangle / N$, is expected to evolve as $u \propto (T_c - T)^{1-\alpha}$. If the regime with $\alpha > 1$ would hold up to the critical temperature, then $u$ would become infinite at $T_c$, which is of course not possible. Figure 13 indeed shows that the value of α deduced from log-log plots of $u$ as a function of $T_c - T$, $\alpha = 1.37$, is close to the estimation obtained from the plot of $c_V$, that is $\alpha = 1.33$. Most importantly, figures 10 to 13 all display a crossover from the first-order regime to another regime in the last few Kelvins below the critical temperature. We checked that the results presented in these figures are converged, that is, they do not vary when the size of the matrix in TI calculations is increased from 4201 to 8201 and the maximum value of $y$ correspondingly increases from about $5000/a$ to about $10000/a$. It still remains, as mentioned in section (3.1), that neither MD simulations nor TI calculations are capable of providing a clear indication of what happens very close to $T_c$. Analogy with the wetting transition [50] and calculations performed with a rougher model [51] suggest that the melting transition is asymptotically second-order, but this point remains to be ascertained.

**5 – Discussion and conclusion**

In this paper, we investigated the properties of a dynamical mesoscopic model of DNA, which shares part of its building blocks with statistical models, in the sense that it is based on site-dependent finite stacking and pairing enthalpies. However, in contrast with statistical models, no explicit temperature dependence is plugged in the dynamical model : instead of site-depend stacking entropies, temperature evolution is indeed governed by the shape of the stacking and pairing interactions. Similarly, the partition function of a loop needs not be estimated and plugged independently, because it is again a direct consequence of the expression of the Hamiltonian. At last, the dynamical model contains nothing like the cooperativity parameter σ. In spite of these differences, we showed that the two models predict denaturation curves and profiles, which are very similar, as can be checked in figures 5 and 6, and which agree with known experimental results.

One could therefore naively believe that these two models provide the same description of the physics of DNA melting. This is unfortunately not exactly the case. As discussed in some detail in [16], statistical models describe DNA denaturation as a phase transition, which order depends on the way the partition function of a loop, and particularly the loop closure exponent $c$, is calculated. When using self-avoiding walks embedded in a three-dimensional space, the exponent $c$ is numerically estimated to be close to $c \approx 1.75$ [52], which corresponds to a second-order phase transition (the boundary between second- and first-order phase transitions is $c = 2$ [10,53]). In contrast, when using loops embedded in chains [19,20], which is probably a better approximation, one gets $c \approx 2.15$ [19,20,54], so that denaturation would correspond to a first-order phase transition, although experimental results can be equally well reproduced using sets of parameters with $c \approx 1.75$ and $c \approx 2.15$ [16]. On the other hand, dynamical models, like the DPB model and the model we propose, suggest that DNA denaturation looks like a first-order phase transition in a rather broad temperature interval, but that this regime is necessarily followed by another regime very close to the critical temperature. The exact nature of DNA melting dynamics in this narrow temperature interval is at present not precisely known. It might correspond to a second-order phase transition, but, as far as we understand it, cannot be excluded that it is a simple crossover between double-stranded and single-stranded DNA.

The order of the denaturation phase transition is a question that interests physicists [11,16,19,20,29,31,50,51,55-60] but is however much less relevant for biologists. What is most important from the biological point of view is probably that statistical models have led to such tools like the MELTSIM and POLAND programs, which allow the rapid estimation of denaturation curves and are helpful for a wide variety of tasks, like for example primer design, DNA control during PCR, and mutation analysis [61]. Because of the length of calculations, dynamical models are of no practical use with that respect. They however usefully demonstrate that different models based on different assumptions lead essentially to the same result, as shown in figures 5 and 6, and that conclusions derived from statistical mechanics calculations are likely not to be model dependent.

Moreover, one of the principal reasons, why thermal denaturation received so much attention, is that the processing of the genetic information stored in the DNA double-helix implies the separation of the two strands, just like thermal denaturation. Although several proteins (helicases, etc…) are usually involved in the separation of the strands during replication or transcription, the question arises whether there exists a relation between genetic maps, i.e. the location of coding regions, and DNA physical stability maps like the one presented in figure 6. If by chance this were the case, then genes could be identified just by looking at thermal disruption maps. First results [34,62,63] seem to indicate that correlation between the two kinds of maps is sometimes perfect but that they are sometimes almost completely uncorrelated. Although this point deserves further attention, dynamical models will again essentially serve as a check to calculations performed with statistical ones, because the answer to the question of the correlation between genetic and thermal stability maps essentially relies on the statistical properties of the sequence.

Statistical models are however often not sufficient to investigate the dynamical or time-dependent properties of more complex systems, like for example the coupled dynamics of DNA breathing and of proteins that selectively bind to single-stranded DNA [64-67]. To this end, one has to develop either a dynamical model or a kinetic one, that is, a model which principal ingredients are reaction probabilities and transfer rates [66,67]. The ratio of calculation times for kinetic and dynamical models being much less unfavourable than for statistical and dynamical ones, dynamical models represent a realistic and interesting alternative to kinetic models for these systems. Moreover, the model for DNA denaturation presented in this paper certainly represents a reliable cornerstone on which one can elaborate the modelling of more complex systems like the interaction of DNA bubbles and of single-strand binding proteins. We are currently working in this direction.

# FIGURE CAPTIONS

**Figure 1**: Plot of the critical force $F_c$, which is required to open a long homogeneous sequence, as a function of the temperature $T$ of the sequence, according to the model of equation (2.1) and the old and new sets of parameters. Solid lines were obtained from equations (2.3), (2.4) and (2.7) and the few open circles from Monte Carlo simulations, as a check to the validity of these equations.

**Figure 2**: Plot, as a function of the length of the sequence $N$, of the difference $T_c - T_c(N)$ between the critical temperatures of an infinitely long homogeneous sequence, $T_c = 359.43$ K (see section 4), and a homogeneous sequence with $N$ base pairs, $T_c(N)$. The dot-dashed line is the least-square fit to the calculated shifts. See [30] for the detail of TI calculations.

**Figure 3**: Plot, as a function of the temperature $T$ of the sequence, of the average base pair separation $\langle y \rangle = \frac{1}{N}\sum \langle y_n \rangle$ for a homogeneous sequence with 1000 bp, obtained from MD simulations (dashed line) and TI calculations (solid line). $\langle y \rangle$ is expressed in Å. The vertical dot-dashed line shows the critical temperature for this sequence ($T_c(N) = 356.73$ K). Each point of the MD curve corresponds to a total accumulation time of 1 μs. See [30] for the detail of TI calculations.

**Figure 4**: Plot, as a function of the temperature $T$ of the sequence, of the average energy in each Morse oscillator and the average total potential energy per base pair, $u = \langle V \rangle / N$, for a homogeneous sequence with 1000 bp, obtained from MD simulations (dashed lines) and TI calculations (solid lines). Energies are expressed in eV. The vertical dot-dashed line shows the critical temperature for this sequence ($T_c(N) = 356.73$ K). Each point of the MD curve corresponds to a total accumulation time of 1 μs. For TI calculations, $u$ was obtained from equation (3.4) and the relation $u = f + T s$. See [30] for the detail of TI calculations.

**Figure 5**: Plot, as a function of the temperature $T$ of the sequence, of the fraction of open base pairs for the 1793 bp human β-actin cDNA sequence (NCB entry code NM_001101) obtained with MELTSIM [17] (dashed line) and the dynamical model proposed in the present paper (solid lines). MELTSIM calculations were performed with the parameters of Blossey and Carlon [16] and a salt concentration $[Na^+]_0 = 75$ mM. Results obtained with three different thresholds ($y_{thresh} = 7$, 10 and 15 Å) are shown for TI calculations performed with the dynamical model. Remember that critical temperatures determined with the $\langle y_n \rangle > y_{thresh}$ criterion are 2 to 3 Kelvins lower than exact ones, as discussed in section 3.2. See [31] for the detail of TI calculations.

**Figure 6**: Plot, as a function of the position of the base pair, of the opening temperature of each base pair of the 1793 bp human β-actin cDNA sequence (NCB entry code NM_001101) obtained with MELTSIM [17] (dashed line) and the dynamical model proposed in the present paper (solid lines). MELTSIM calculations were performed with the parameters of Blossey and Carlon [16] and a salt concentration $[Na^+]_0 = 75$ mM. Results obtained with three different thresholds ($y_{thresh} = 7$, 10 and 15 Å) are shown for TI calculations performed with the dynamical model. Remember that critical temperatures determined with the $\langle y_n \rangle > y_{thresh}$ criterion are 2 to 3 Kelvins lower than exact ones, as discussed in section 3.2. See [31] for the detail of TI calculations.

**Figure 7**: Plot of the specific heat per particle, $c_V$, as a function of temperature $T$ for the 1793 bp human β-actin cDNA sequence (NCB entry code NM_001101), obtained from TI calculations performed with the dynamical model proposed in this paper. $c_V$ is expressed in units of the Boltzmann constant $k_B$. See [31] for the detail of TI calculations.

**Figure 8**: Plot, as a function of the temperature $T$ of the sequence, of the entropy per base pair, $s$, for an infinitely long homogeneous sequence and sequences with $N = 1000$ and $N = 100$ bp. These results were obtained from TI calculations performed as described in [29,30]. $s$ is expressed in units of the Boltzmann constant $k_B$. The solid curve for the infinitely long chain is interrupted in the temperature interval where the TI method is not valid.

**Figure 9**: Plot, as a function of the temperature $T$ of the sequence, of the specific heat per base pair, $c_V$, for an infinitely long homogeneous sequence and sequences with $N = 1000$ and $N = 100$ bp. These results were obtained from TI calculations performed as described in [29,30]. $c_V$ is expressed in units of the Boltzmann constant $k_B$. The solid curve for the infinitely long chain is interrupted in the temperature interval where the TI method is not valid.

**Figure 10**: Log-log plot, as a function of the temperature gap $T_c(N) - T$, of the specific heat per base pair, $c_V$, for an infinitely long homogeneous sequence and sequences with $N = 1000$ and $N = 100$ bp. These results were obtained from TI calculations performed as described in [29,30]. $c_V$ is expressed in units of the Boltzmann constant $k_B$. The dot-dashed straight line shows the slope corresponding to a critical exponent $\alpha = 1.33$.

**Figure 11**: Log-log plot, as a function of the temperature gap $T_c(N)-T$, of the average base pair separation, $\langle y \rangle$, for an infinitely long homogeneous sequence and sequences with $N=1000$ and $N=100$ bp. These results were obtained from TI calculations performed as described in [29,30]. $\langle y \rangle$ is expressed in Å. The dot-dashed straight line shows the slope corresponding to a critical exponent $\beta = -1.41$.

**Figure 12**: Log-log plot, as a function of the temperature gap $T_c - T$, of the correlation length, $\xi$, for an infinitely long homogeneous sequence. This result was obtained from TI calculations performed as described in [29]. $\xi$ is expressed in units of a length characterizing the separation between successive base pairs. The dot-dashed straight line shows the slope corresponding to a critical exponent $\nu = 1.47$.

**Figure 13**: Log-log plot, as a function of the temperature gap $T_c(N)-T$, of the average potential energy per base pair, $u = \langle V \rangle / N$, for an infinitely long homogeneous sequence and sequences with $N=1000$ and $N=100$ bp. These results were obtained from TI calculations performed as described in [29,30]. $u$ is expressed in eV. The dot-dashed straight line shows the slope corresponding to a critical exponent $1-\alpha = -0.37$.

FIGURE 1

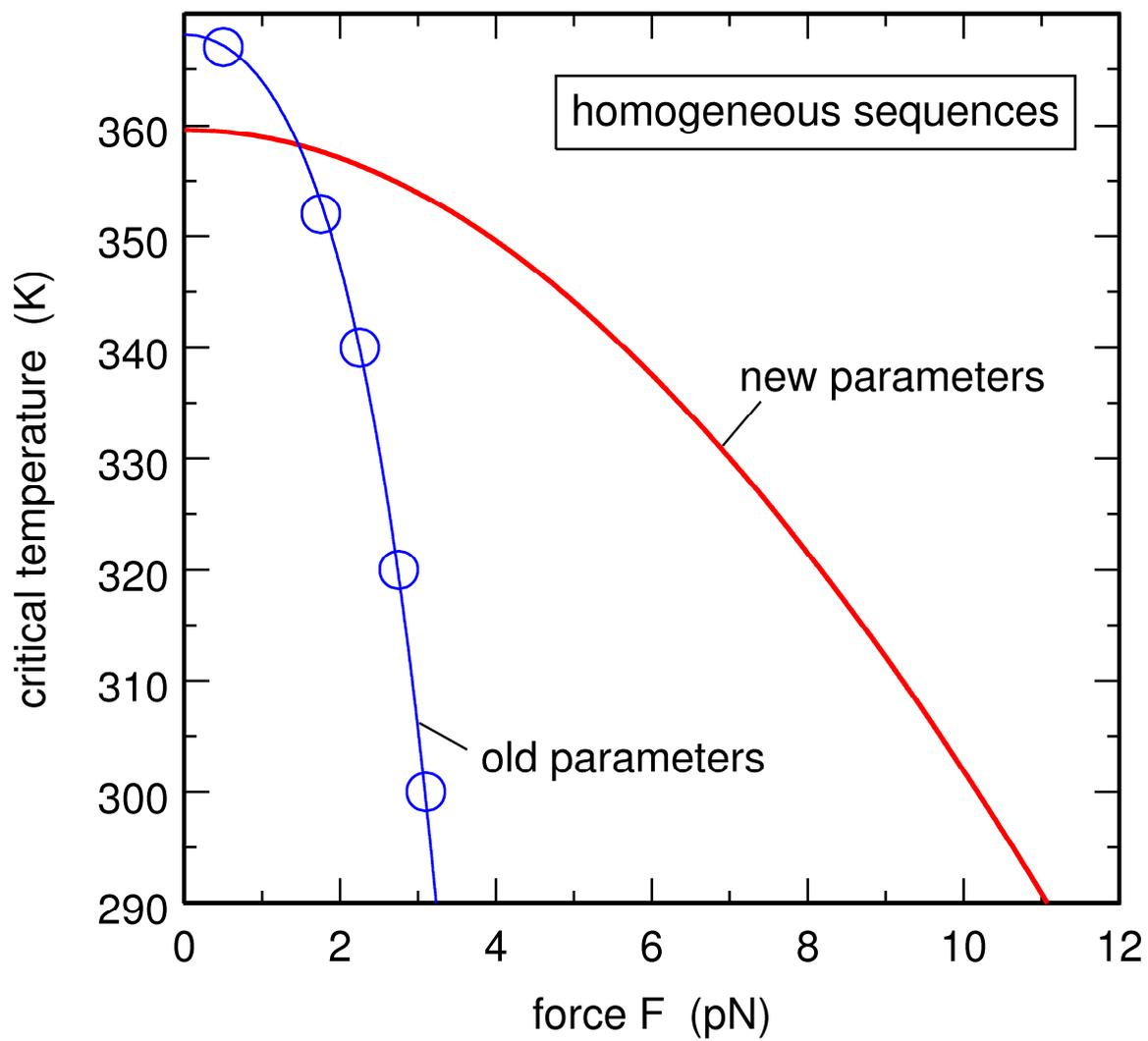

FIGURE 2

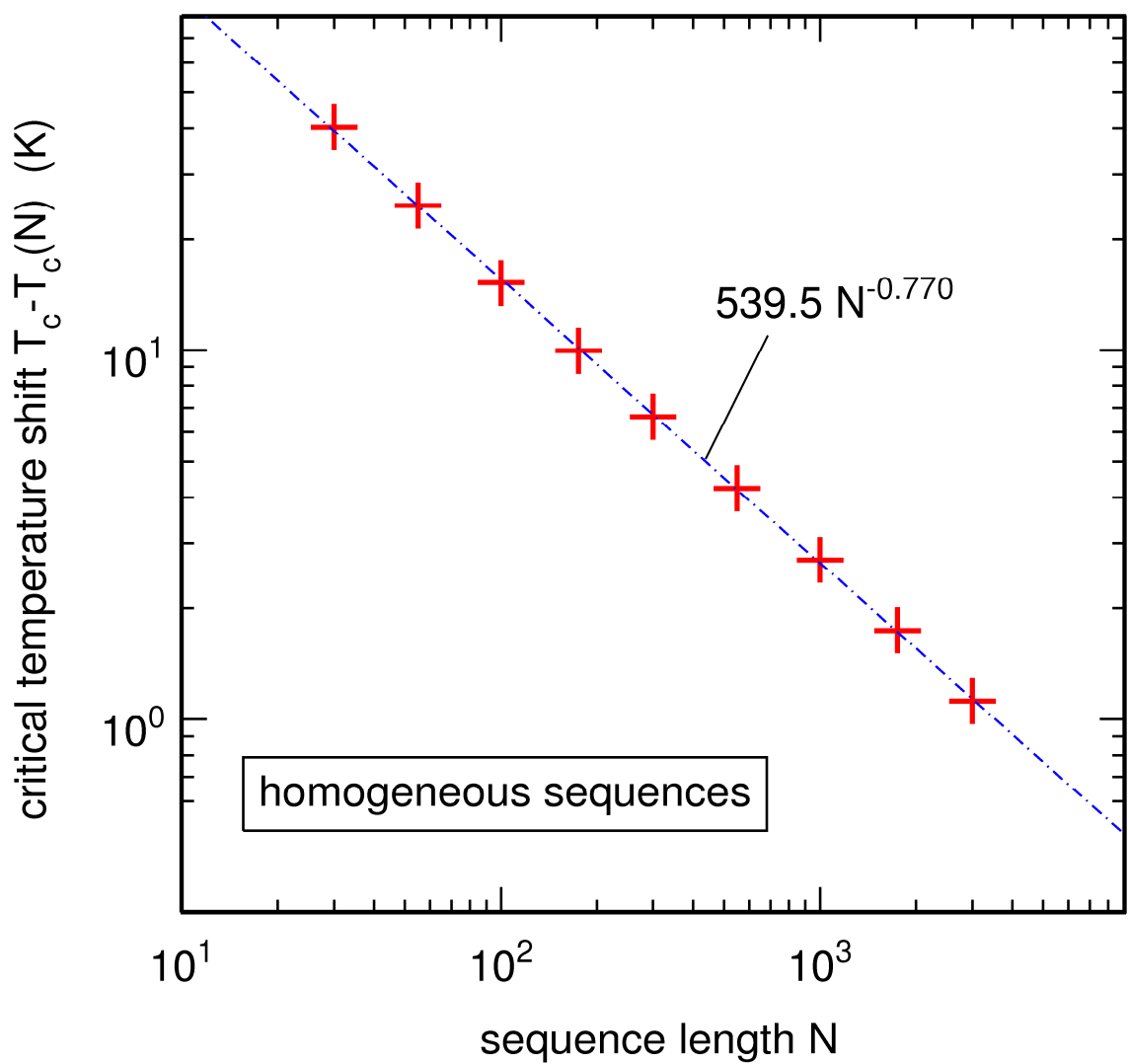

$539.5\ N^{-0.770}$

homogeneous sequences

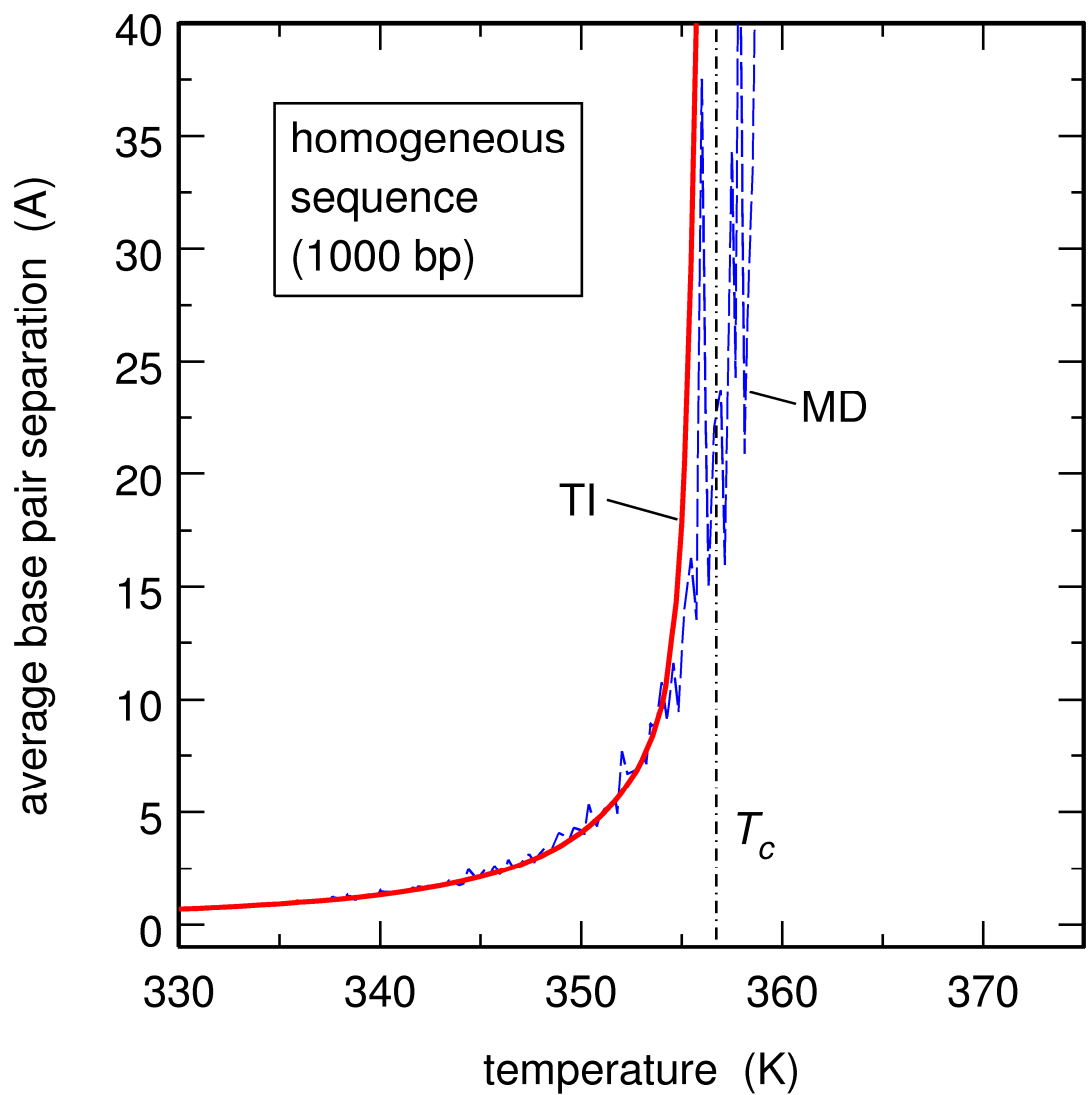

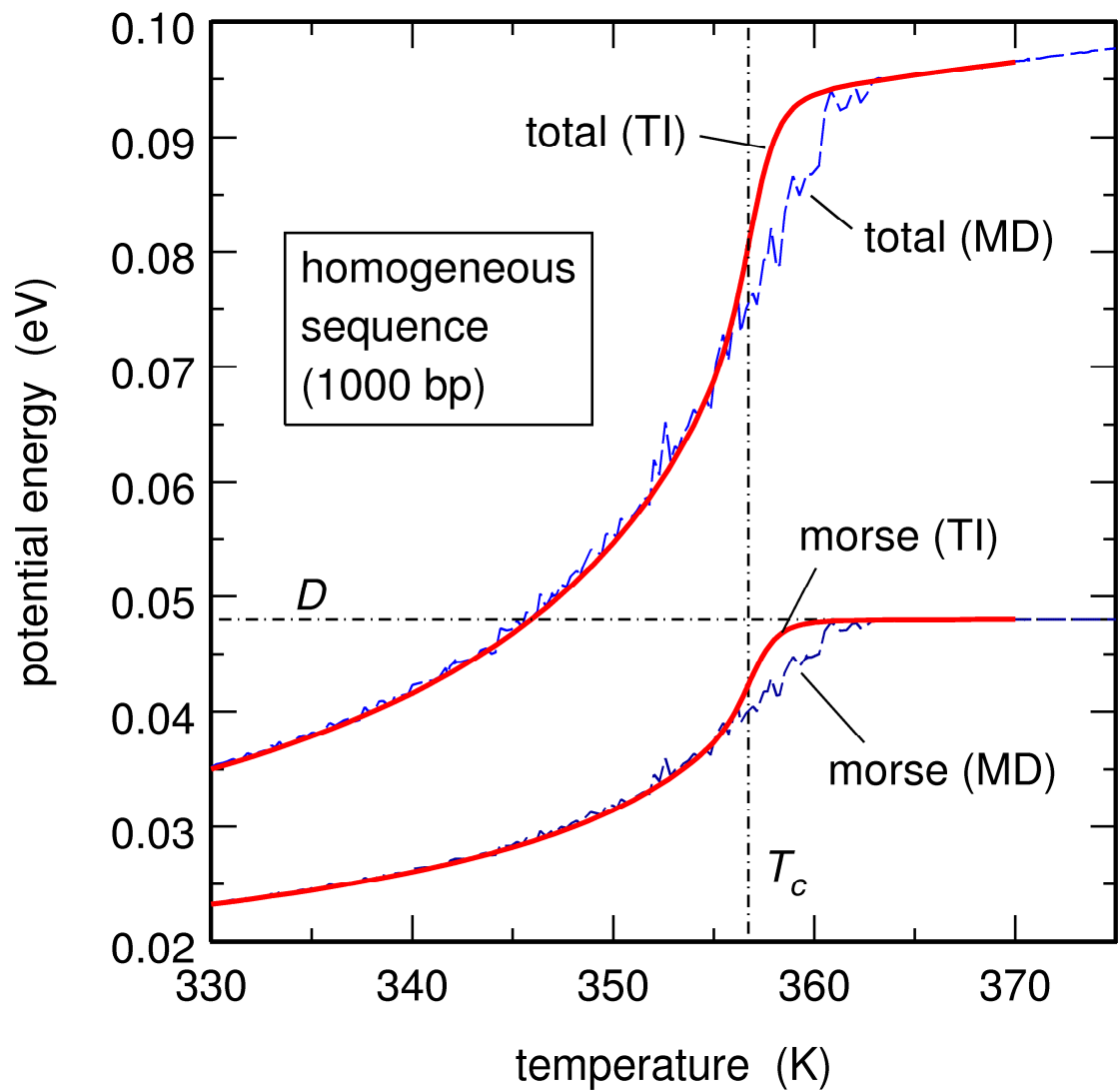

FIGURE 5

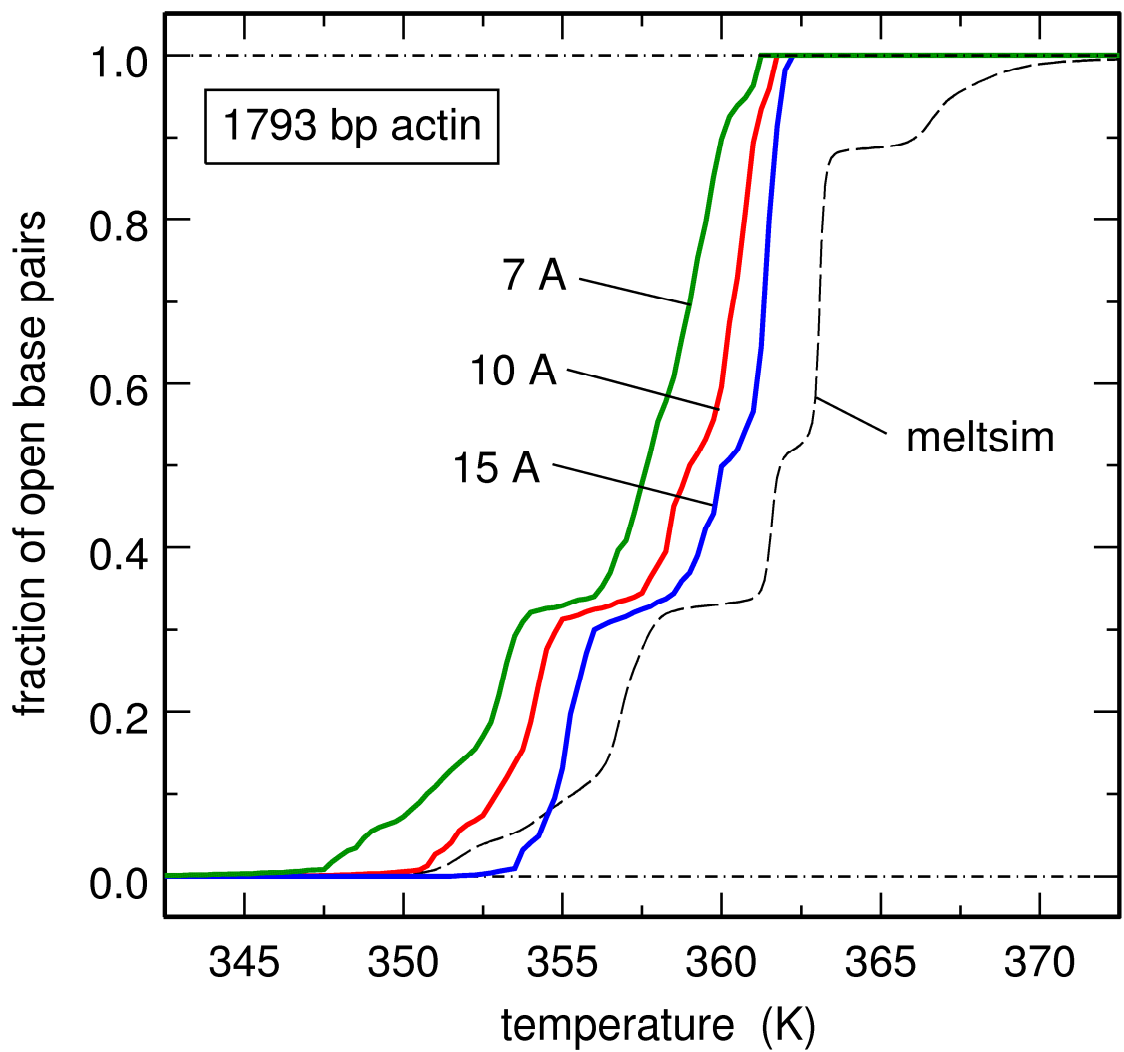

FIGURE 6

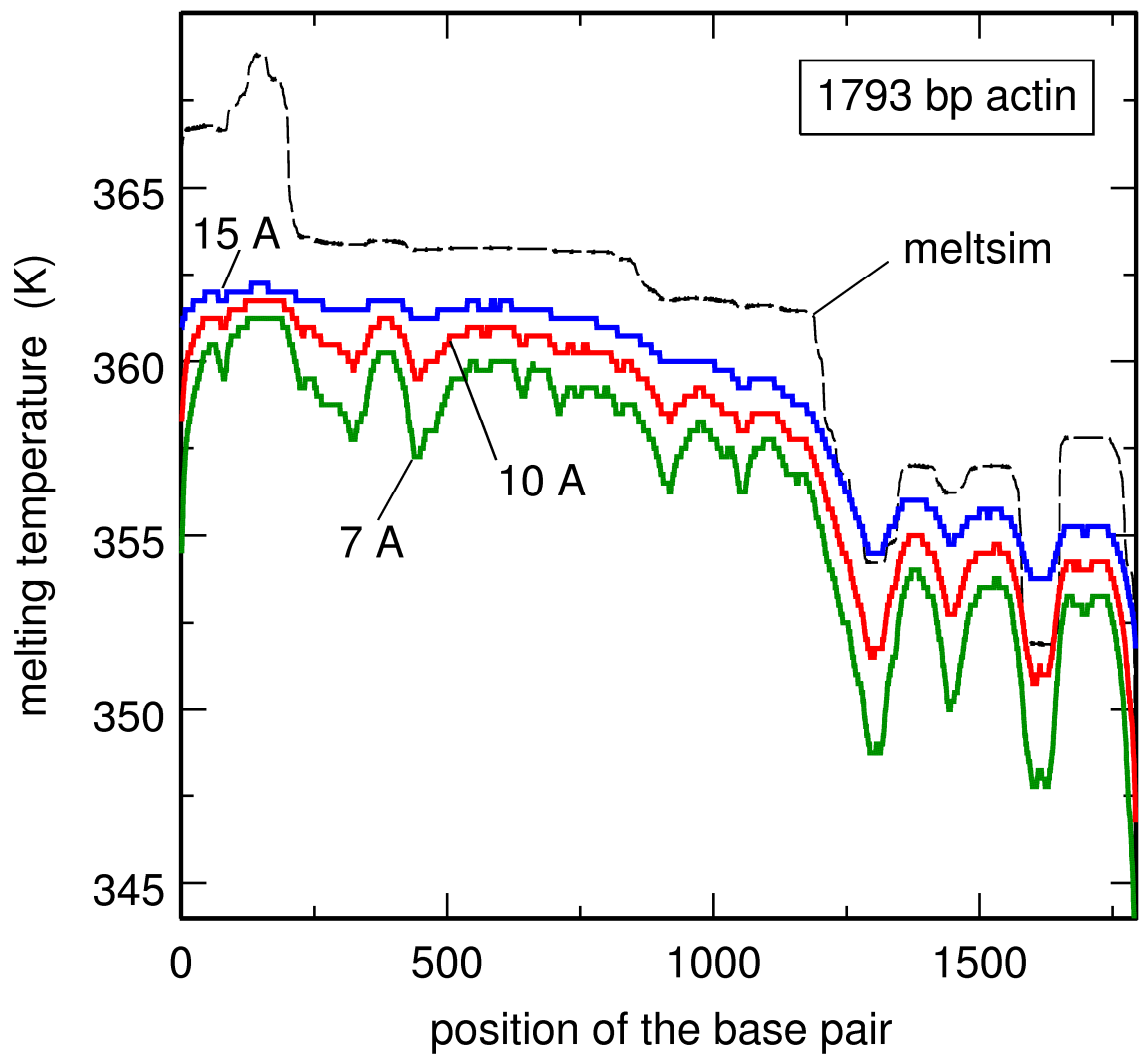

FIGURE 7

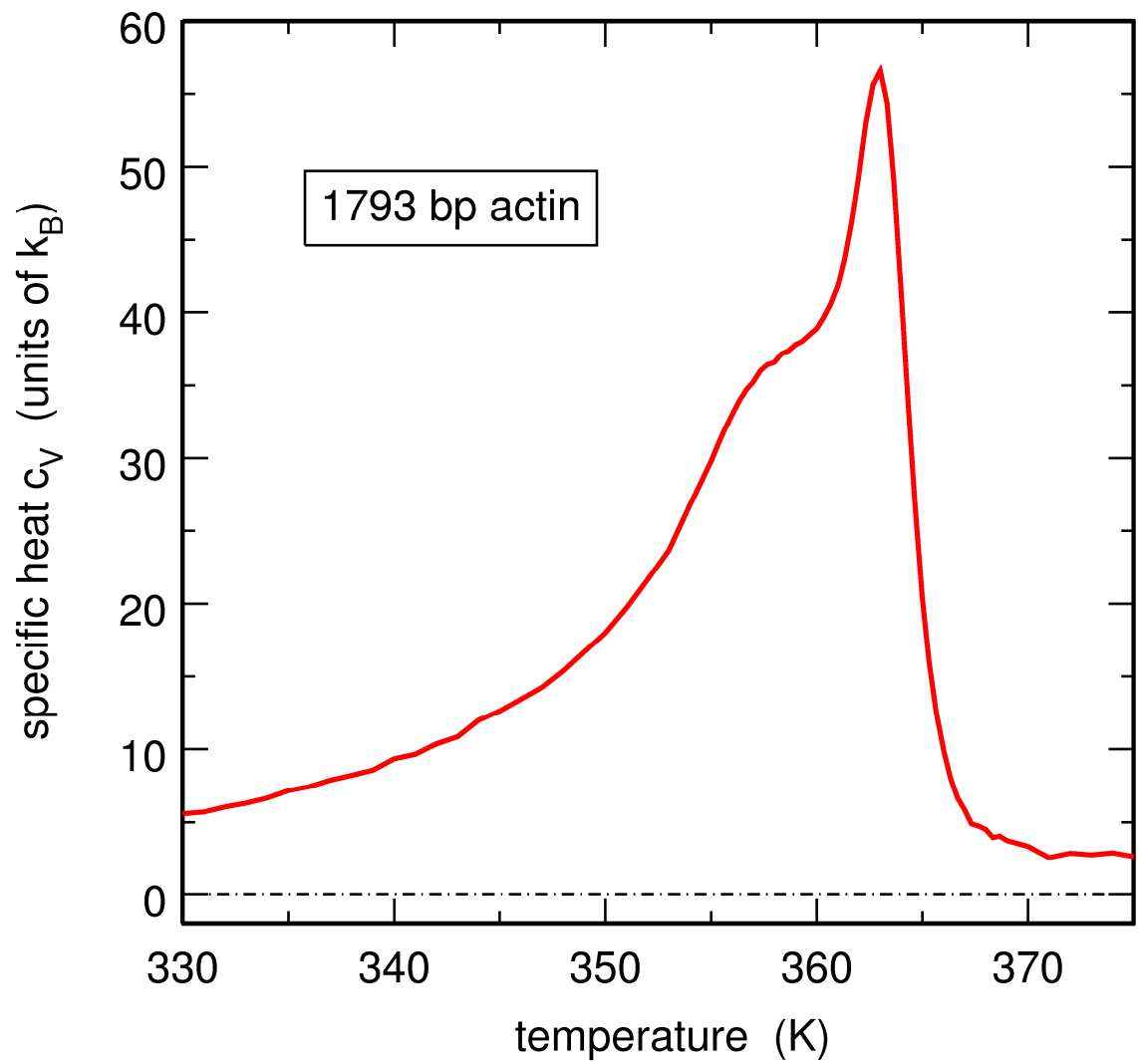

FIGURE 8

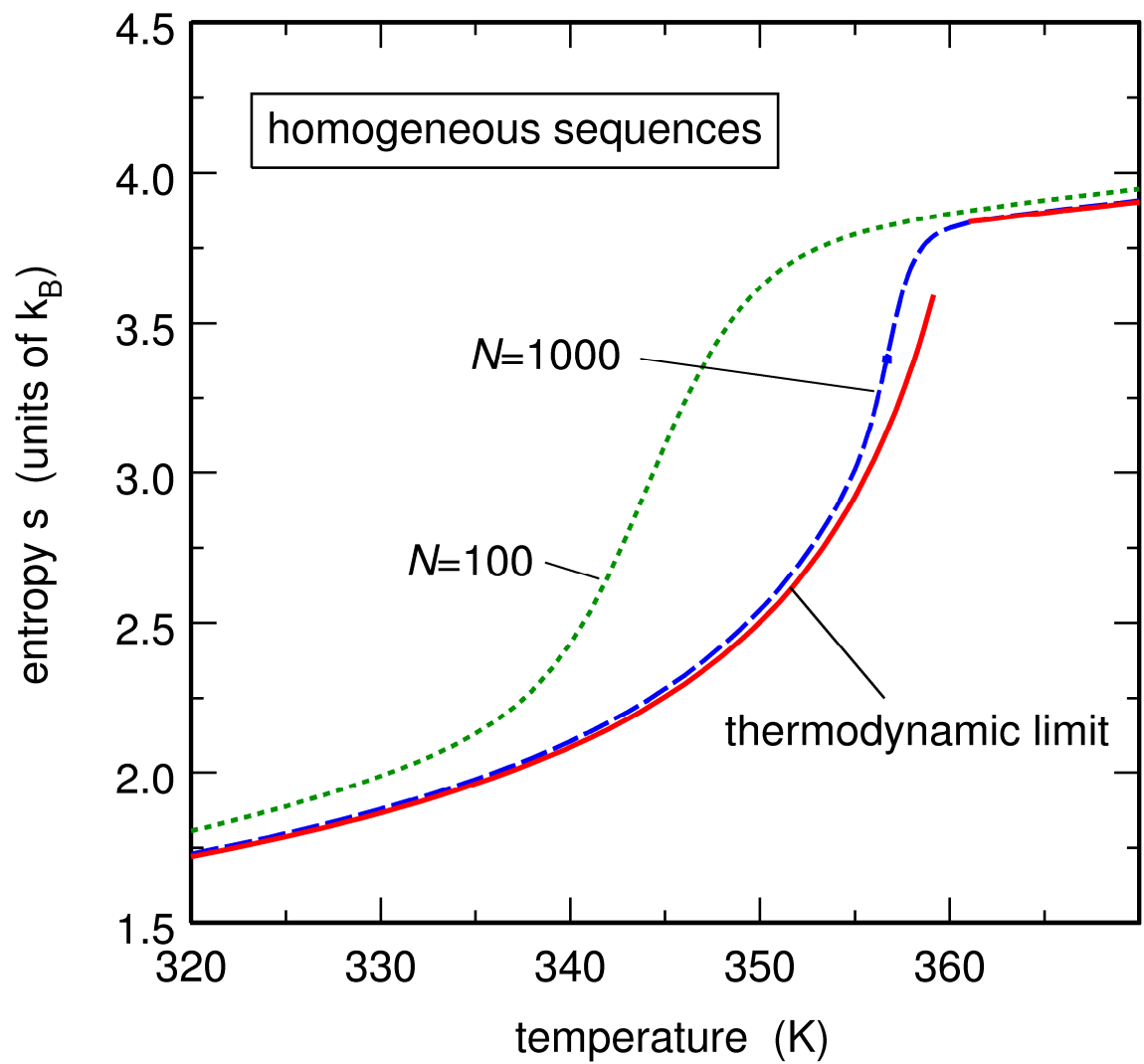

FIGURE 9

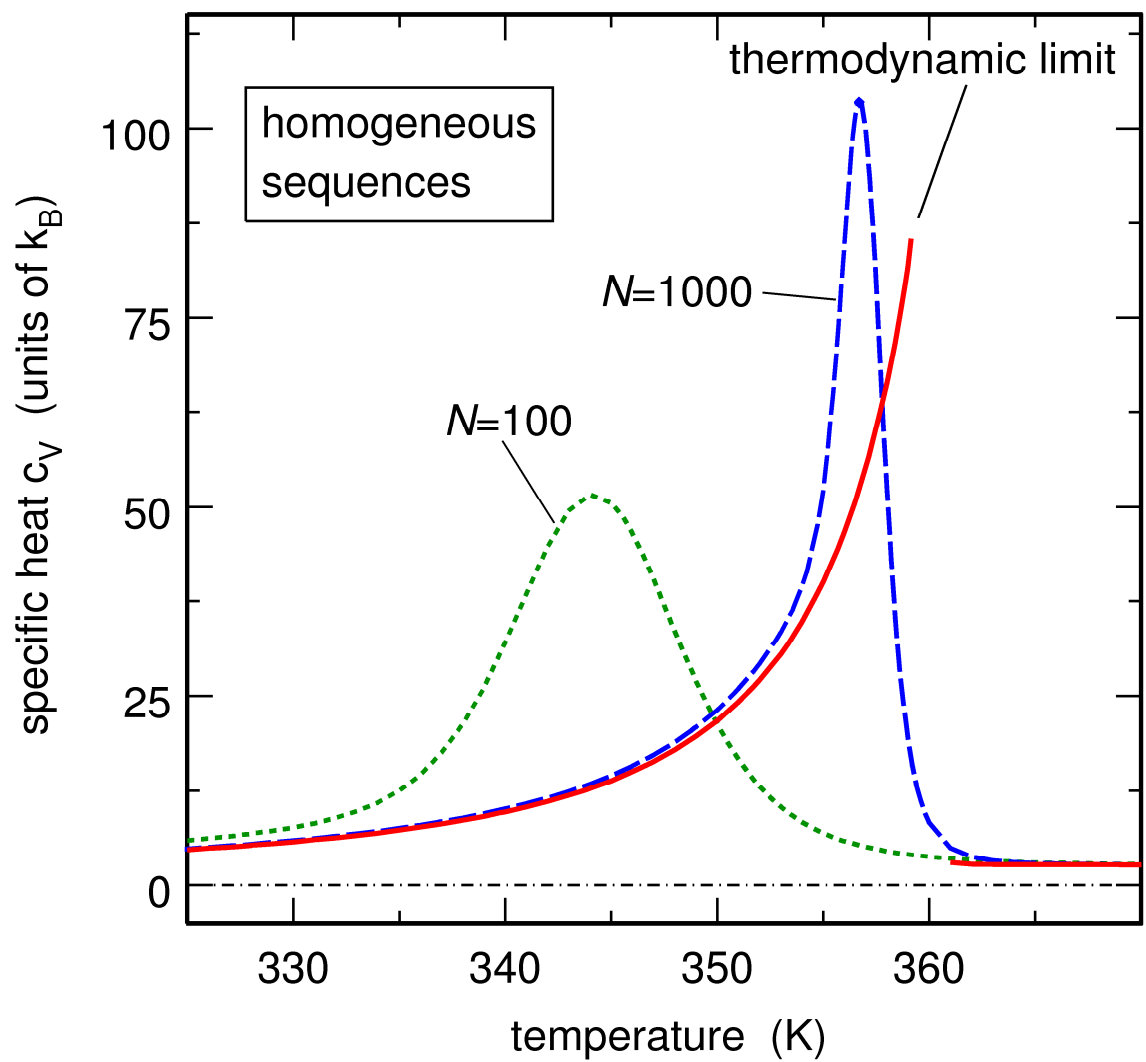

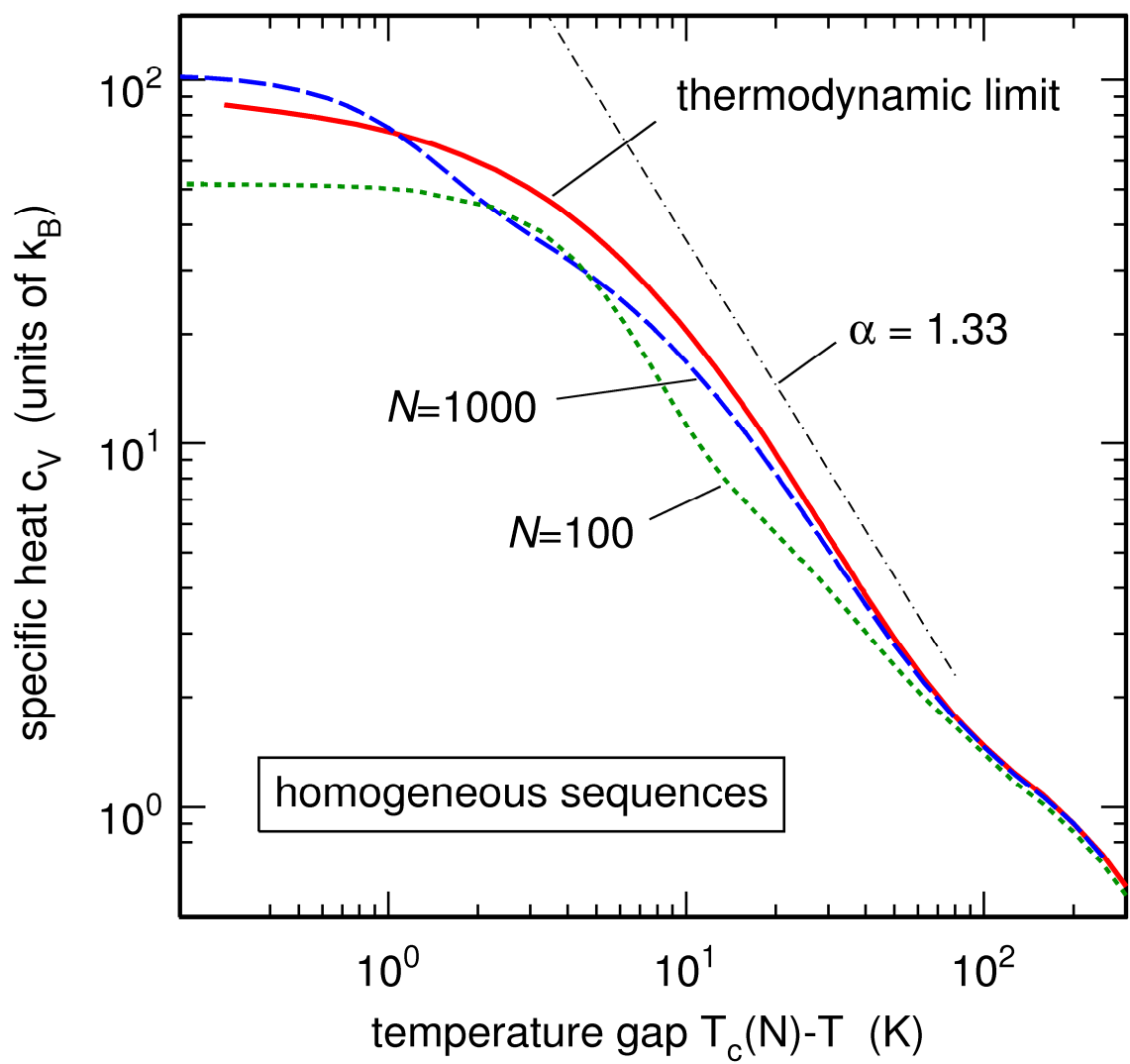

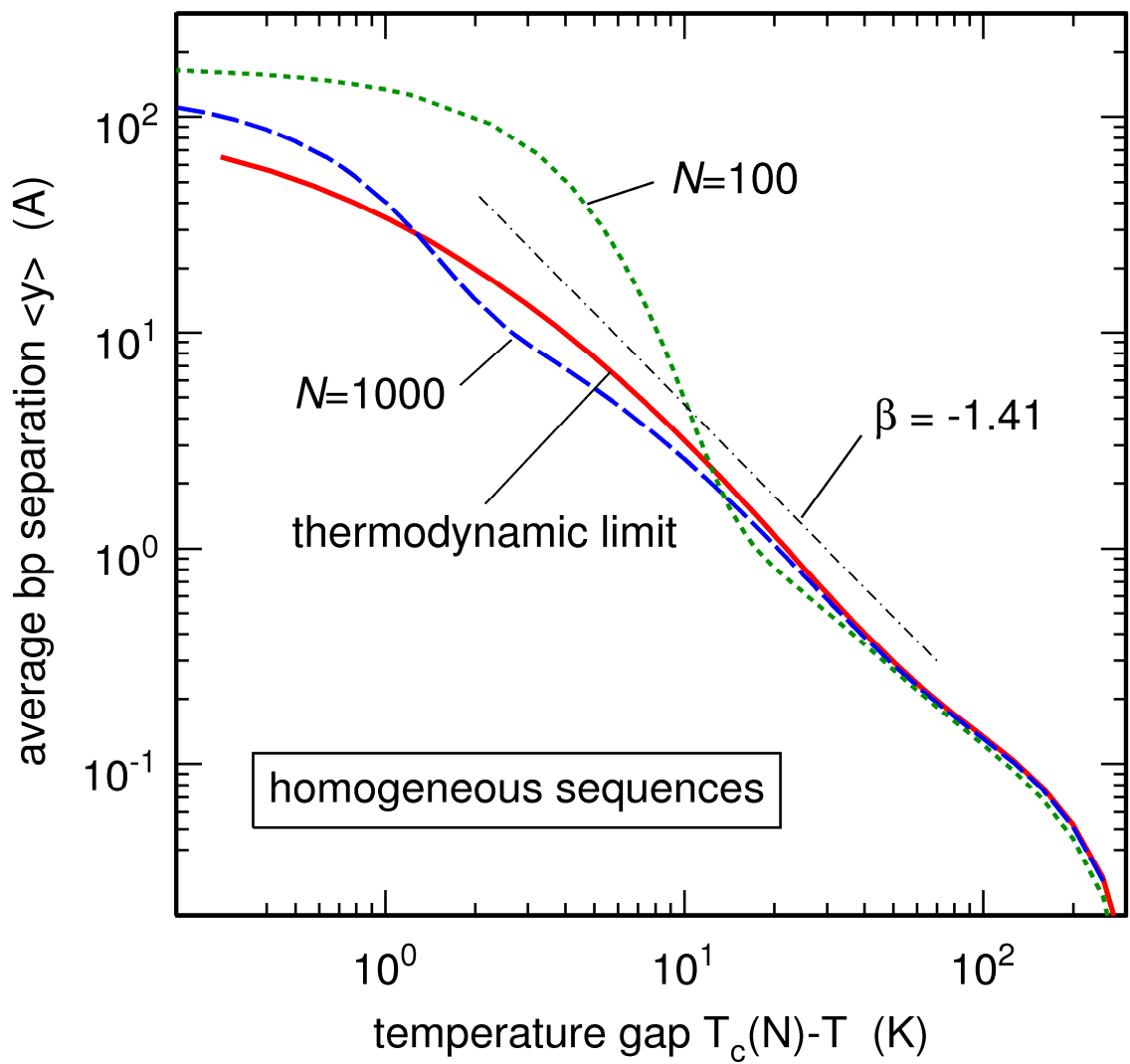

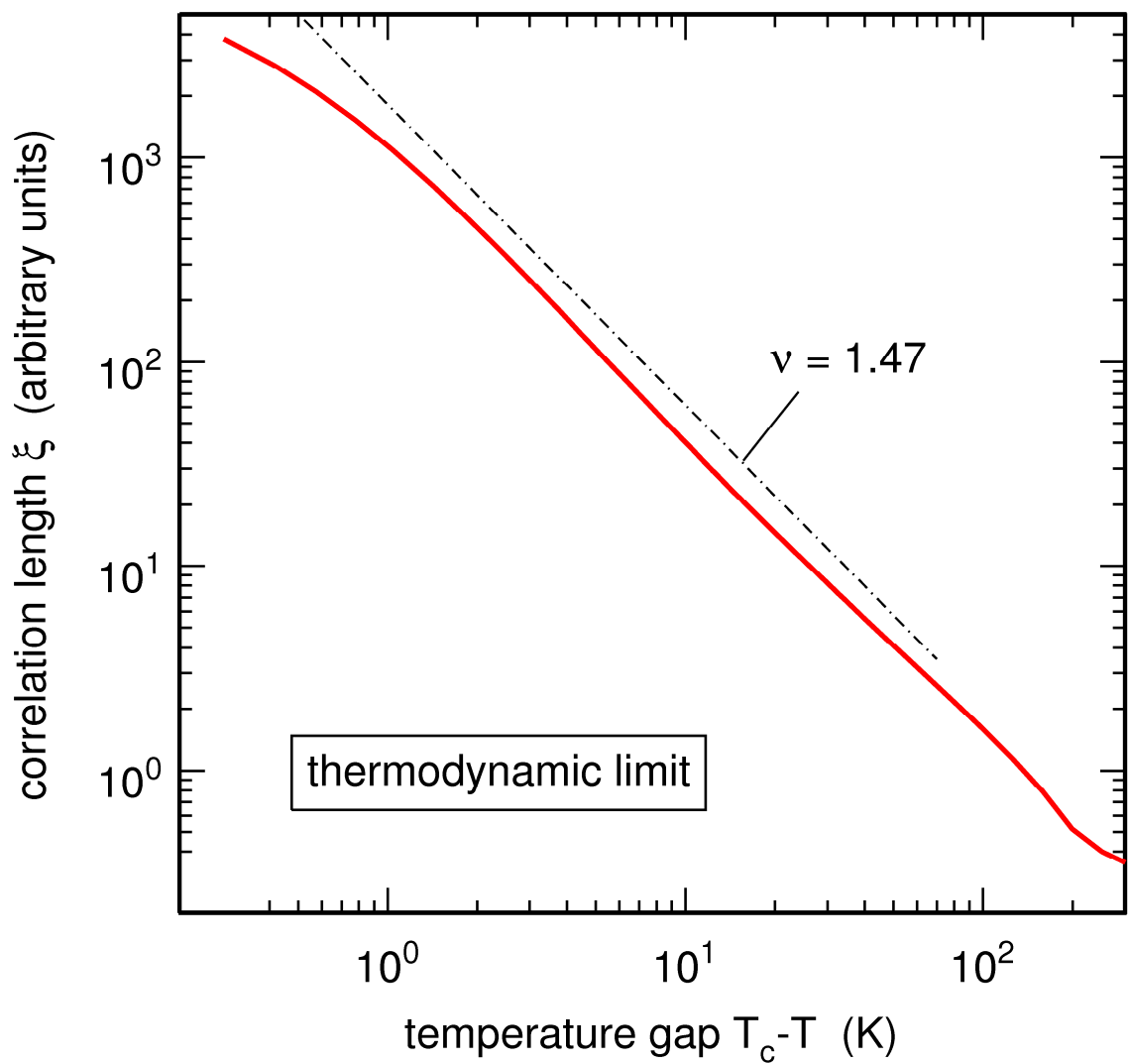

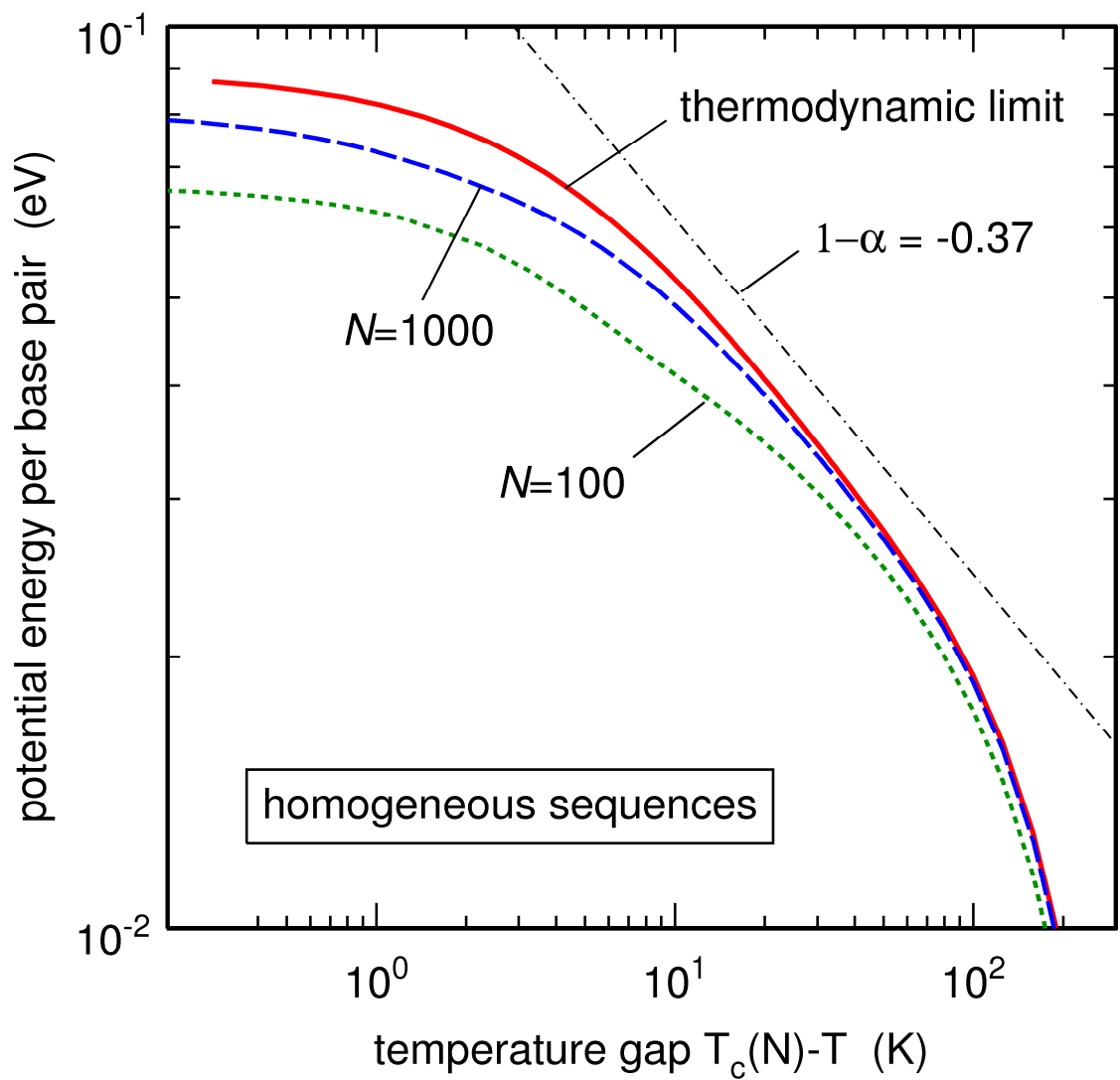